\documentstyle[12pt,aaspp,epsf,epsfig]{article}
\def\lsim{\lower.5ex\hbox{$\; \buildrel < \over \sim \;$}}
\def\gsim{\lower.5ex\hbox{$\; \buildrel > \over \sim \;$}}
\def\he#1{\hbox{$^{#1}{\rm He}$}}

\def\li#1{\hbox{$^{#1}{\rm Li}$}}
\def\b1#1{\hbox{$^{1#1}{\rm B}$}}
\def\be#1{\hbox{$^{#1}{\rm Be}$}}
\def\msun{${\,M_\odot}$}
\def\etal{{\it et al.}~}
\def\beginapjbib{\begingroup \section*{\large \bf References}
   \parskip=.5ex plus 1.0pt
   \def\bibitem{\par \noindent \hangindent\parindent
      \hangafter=1}}
\def\endapjbib{\par \endgroup}

\def\beq{\begin{equation}}
\def\eeq{\end{equation}}
\begin{document}
\rightline{IAP-nov 1998}
\rightline{astro-ph/9811327}
\rightline{November 1998}
%\tighten
\title{Lithium-6 : Evolution from Big Bang to Present}

\author{Elisabeth Vangioni-Flam}
\affil{Institut d'Astrophysique de Paris\\ 98bis
Boulevard Arago\\ 75014 Paris, France}

\author{Michel Cass\'e}
\affil{Service d'Astrophysique\\DSM, DAPNIA, CEA\\
 Orme des Merisiers 91191 Gif sur Yvette, France\\ 
and \\ Institut d'Astrophysique de Paris\\98bis Boulevard Arago\\75014
 Paris, France}
 
\author {Roger Cayrel}
\affil {Observatoire de Paris\\ 61 avenue de l'Observatoire\\
  75014 Paris, France}

\author {Jean Audouze}
\affil {Institut d'Astrophysique de Paris\\ 98 bis Bd Arago\\
  75014 Paris, France}

\author {Monique Spite}
\affil {Observatoire de Paris-Meudon\\ 92125 Meudon, CEDEX France}

\and
 
\author {Fran\c cois Spite}
\affil {Observatoire de Paris-Meudon\\ 92125 Meudon, CEDEX France}

\begin{abstract}

The primordial abundances of Deuterium, \he4, and 
\li7 are crucial to determination of the baryon density of the 
Universe in the framework of standard Big Bang 
nucleosynthesis (BBN). \li6 which is only 
produced in tiny quantities and it is generally not considered to be 
a cosmological probe. However, recent major observational 
advances have produced an estimate of the \li6/\li7 ratio 
in a few very old stars in the galactic halo which impacts the
question  whether or not
the  lithium isotopes are depleted in the outer layers of halo stars, 
through proton induced reactions at the base of (or below) the convective
zone.
Here, we use i) an
empirical relation, independent of any evolutionary model, to set an 
upper limit on the \li6 rise compatible with the very existence 
of the Spite's plateau (i.e. the flat lithium abundance measured in 
very old stars of the halo of our Galaxy of different iron content) 
and ii) a well founded evolutionary model of light elements based 
on spallation production (Vangioni-Flam \etal 1997, 1998).

  Indeed, \li6 is
 a pure product of spallation through the major production reactions, fast
 oxygen and alphas interacting on interstellar H, He (especially in
 the early Galaxy). The 
 rapid nuclei are both synthesized and accelerated by SN II. In this context,
 the \li6 evolution
 should go in step with that of beryllium and boron, recently observed by the
 Keck and HST telescopes. \li6 adds a new constraint on the early spallation in
 the
 Galaxy. In particular, if confirmed, the $\li6$/$\be9$ ratio observed in
 two halo stars (HD 84937, BD +$26 \deg3578$ = HD 338529) gives strong
 boundary conditions
 on the
 composition and the spectrum of the rapid particles involved.   
 
 Both methods converge to show  that \li6 is essentially intact in halo stars,
  and 
a fortiori \li7, which is more tightly bound. Moreover, 
extrapolating empirical and theoretical evolutionary 
curves to the very low metallicities, we can define a range of the 
 \li6 abundance in the very early Galaxy consistent with Big Bang
 nucleosynthesis (5.6 $10^{-14}$ to 3. $10^{-13}$) . Following the
 evolution at increasing metallicity, we explain the abundance in the
 solar system within a factor of about 2.
 The whole evolution 
 from Big Bang to present is reasonably reproduced, which demonstrates
 the general consistency of the present analysis of \li6.
 
 The baryonic density derived from both 
lithium isotopes is between 1.5 to 3.5 per cent of the critical one,
 in good agreement 
with the determination based on independent analyses.
Consequently, thanks to these new data and theoretical developments, 
we show that \li6 can be used to establish stellar \li7 abundances as a 
 valid tracer due to the 
fact that it allows to reinforce the Spite's plateau as a primordial
 test of BBN; on the other hand, its early evolution can be used 
 to corroborate the calculated BBN abundances. In the framework of this work,
 a pregalactic $\alpha$ + $\alpha$ process producing \li6 is not necessary. 
 Finally, thanks to \li6, the
 physics of spallative production of light elements should be more easily
 mastered when more data will become available.

\end{abstract}

\keywords{Cosmic-rays, Galaxy :abundances, Nucleosynthesis, Big Bang\\
Nucleosynthesis}

\newpage

\section{Introduction}

{\ } \li6 is a rare and fragile isotope. Indeed, this nucleus is 
destroyed in stellar interiors at temperatures higher than $2 \times
10^6$ K. In hot (low mass) stars composing the Spite's plateau,
  according to the depth of the external convection 
 this isotope could survive partially
or totally in stellar atmospheres  (where they are observed) even though
this represents at most a few percents of the stellar mass. The
preservation of the stellar surface layer in these stars has been well
established by the observation of the (more durable) \li7 isotope
(Spite \& Spite, 1982).  It is well known that the
lithium isotope is found to have a nearly constant abundance independent
of metallicity (for [Fe/H]$< -1.3$) and temperature (for surface
temperatures $ \gsim 5500$ K). The lithium plateau is generally regarded
to represent the primordial abundance of \li7 and is of key importance to
understanding Big Bang Nucleosynthesis (see e.g. Fields \etal 1996,
 Schramm and Turner 1998).
Unlike \li7, \li6 is not produced in large quantities in the Big Bang
(\li6/\li7 $\sim$ 0.001, see e.g. Figure 1),
but rather by non thermal reactions (spallation) induced by
energetic collisions of fast nuclei with the interstellar medium (ISM). 

There is of course a nagging uncertainty as to whether or not the
observed \li7 abundance in the plateau stars is in fact
representative of the primordial abundance or partially
depleted by non-standard stellar processes. 
%Deliyannis and Malaney (1995) 
%have suggested a possible production by stellar flares. This non  thermal
%mode of stellar synthesis, however, has proved to be  unlikely (Lemoine
%\etal 1998). 
 The discovery by Spite and Spite (1982) of a quasi constant 
Li abundance, independent of the stellar metallicity, now confirmed
 down to very low 
values (about [Fe/H] = -4) is a landmark of the BBN theory. The 
observed Li/H ratio of about $1.6 \times 10^{-10}$ has been widely
 confirmed (Spite 1997, 
Bonifacio and Molaro 1997) and taken at face value is concordant 
with standard BBN calculations. 
However, some stellar models originally aimed at 
explaining the paucity of lithium in the sun and other Pop I stars, 
extrapolated to low metallicity, predict a significant depletion of 
\li7 in the external layers of halo stars (Deliyannis \etal 1990, 
Charbonnel \etal 1992) due to burning at the base of the convective 
envelope. The depletion factor is model dependent and open to 
debate (e.g. Pinsonneault \etal 1992, 1998, Chaboyer 1998, Vauclair and
 Charbonnel
1998). The fates of \li6 and \li7 are linked in the atmospheres of 
stars, but \li6 has a lower destruction temperature than 
\li7 ( $2.0 \times 10^{6}$ K versus $2.5 \times 10^{6}$ K). Thus, the
observation of \li6 in stellar atmospheres provides a very strong
constraint on the depletion factors and consequently on the effective
primordial  values of the lithium isotopes.

Recently, major observational advances (Smith \etal 1993, 
Hobbs and Thorburn 1994, 1997) have yielded an estimate of the 
\li6/\li7 ratio in a few very old stars in the halo of the 
Milky Way.
Specifically, the lithium isotopic ratio has been measured in HD 84937,
most  recently by Cayrel \etal (1998) with a very good S/N ratio finding
\li6/\li7 = 0.052 $\pm$ 0.015  at [Fe/H] = -2.3, and by Smith \etal (1998)
who found \li6/Li = 0.06 $\pm$ 0.03 at [Fe/H] = -2.2.
  Smith \etal (1998) have reported
one other positive detection in BD +$26\deg 3578$ (=HD 338529) with
 \li6/Li = 0.05
$\pm$ 0.04 at [Fe/H] = -2.3. For the other halo stars  examined,
only upper limits are available. This is not surprising since
 i) above [Fe/H] = -1, the convection zone deepens and \li6 reaches 
high temperatures and is therefore burnt, ii) at very low metallicity, 
the \li6/\li7 ratio is so small that it becomes heroic to try 
to separate the two lithium isotopes. More observations are 
required around [Fe/H] = -2 to improve the situation.

Provided the galactic \li6 evolution can be followed through its
 nucleosynthesic mechanisms, these detections permit one 
 to draw important conclusions
on the depletion factor of both lithium isotopes in the stars of interest. 
 Limits on the depletion factors allow in turn to get
 constraints on their primordial  abundances.

In fact, non thermal lithium production occurs through 
well identified spallation and fusion processes (Read and Viola 1984)
          mainly fast O and alphas impinging 
 on interstellar H and He, specifically  in the early Galaxy
 (Vangioni-Flam et al 1998).
 Here, we study the evolution of 
the cumulated abundance of \li6 in the ISM  i) 
through a galactic evolutionary model (Vangioni-Flam \etal 1996, 1997, 1998),
and ii) through an empirical (model independent) criterion based on the
 requirement that \li6 evolution vs [Fe/H] is linear at least from [Fe/H]
= -2.5 up to -1. In principle, 
the comparison between the calculated value of the \li6/H ratio in 
the ISM and that observed in the stellar envelope at a measured 
[Fe/H] leads to an estimate of the depletion factor of this isotope.
This provides important constraints at the intersection of 
three different astrophysical domains, stellar evolution, non 
thermal nucleosynthesis and cosmological nucleosynthesis.
 In section 2 we develop the BBN and spallation production 
mechanisms of \li6. In section 3, we study the general evolution of \li6
 using  empirical and theoretical 
constraints and we derive i) the depletion factor of this isotope in stellar 
envelopes and  ii) the range of \li6 abundance in the very early galaxy to 
 be compared to 
the primordial \li6 abundance in relation to  \li7 from BBN.
 In section 4, we present 
the astrophysical and cosmological consequences of the \li6 observations.

\section{ Production of \li6 in the Big Bang and by 
spallation}

\subsection{Big Bang nucleosynthesis}

The Big Bang production of \li6 is dominated by the 
D $(\alpha, \gamma)$ \li6 reaction (Thomas \etal 1993, Schramm 1992,
1994, Nollet \etal 1997). No direct measurement of the cross section
  of this reaction 
has been performed below 1 MeV. However, the Coulomb breakup 
technique (Kiener \etal 1991) provides an indirect estimate which is
  in qualitative
 agreement with the theoretical extrapolation at low energy of Mohr \etal
 (1994).
Recently, the European Collaboration between nuclear 
physicits and astrophysicits led by Marcel Arnould (NACRE : 
European Astrophysical Compilation of Reaction Rates) has 
delivered a consistent compilation of thermonuclear reaction rates 
of astrophysical interest, among them is the D  $(\alpha, \gamma)$ \li6 
reaction (Angulo \etal 1998). They conclude that the reaction rate 
based on the Mohr \etal (1994) $S$ factor is the most relevant. This 
rate is similar to that of Caughlan and Fowler (1988), in the 
temperature range of cosmological interest. The two estimates 
agreee to within a few percents.
%Nollet \etal (1997) have considered all the published 
%evaluations of this reaction rate. However, most of the 
%extrapolations at low temperature depart considerably from the 
%Kiener \etal (1991) estimate except that of Mohr \etal (1994).  
Following the recommendation of Kiener (1998, private 
communication) and Angulo \etal (1998) we adopt the Mohr estimate.
 Note that the upper 
limit given by Cecil \etal (1996) is much higher. This upper limit is 
indeed related to the bad sensitivity of the detector used (Kiener, 
private communication). 
Figure 1 shows the variation of the \li6 and \li7 
abundances versus the baryon-to-photon ratio, $\eta$, calculated with 
our BBN model. Also shown  are the \be9, \b10 and \b11  
abundances calculated with updated reactions rates, including the new 
  \b10$(p,\alpha)$\be7 reaction ( adopted from  Rauscher and
Raimann, 1997). Clearly,  the calculated
primordial Be and B abundances are negligibly small, much more so
 than \li6 and the observed abundances  were not produced in BBN
(Delbourgo-Salvador and Vangioni-Flam 1992).

\subsection{Spallation production}

Aside from its marginal Big Bang origin, \li6 is a product of 
spallation operating through the $\alpha + \alpha$  reaction and the
collisional break  up of C, N,  O nuclei. Spallation agents are i) galactic
cosmic rays  (GCR), specifically acting in the galactic disk through
p,$\alpha$ + He,  CNO $\rightarrow$ \li6, \li7 and ii) fast  nuclei ($\alpha$,
 C,O)
produced and  accelerated by SN II and  fragmenting on H and He in the
ISM,  efficient in the halo phase as recently discussed by Vangioni-Flam \etal
(1998). 
This Low Energy Component (LEC), distinct from standard GCR, is
  thought to be necessary due
 to an observed  linear  relationship between Be, B and [Fe/H], in metal poor
halo stars  (Duncan \etal 1992, 1997, Boesgaard  
and King 1993, Molaro \etal 1995, 1997, Garc\'ia-L\'opez \etal
 1998 and references therein). Thus, the production rate is independent of the 
ISM metallicity , this is the definition of a primary product.
  \li6 itself is a primary
 product in both components due to the $\alpha$ + $\alpha$ reactions, though 
 less efficient in the GCR case since the cross section is peaked around 
 10 MeV/n which is low compared to their average energy ( about 1 GeV/n),
 whereas the
 mean energy of the LEC component corresponds to the maximum of the
 $\alpha$ + $\alpha$ cross section. Specifically, the ratio of the  \li6
 production cross section averaged over the spectrum is about 50 times
 higher for the LEC than for 
 GCR  (using the spectrum given in Lemoine et al 1998). In figure 2
 which presents the \li6/H evolution vs [Fe/H] (see below) it can
 be seen that the cumulated \li6 abundance is overwhelmed by the LEC, though the
 same nuclei are involved in the early Galaxy. Clearly the difference is 
 due to the distinct  spectral shapes.

 In contrast, standard GCR predicts that BeB production rates depend 
 both on the CNO abundance in the ISM at a given time and on the intensity 
 of the cosmic-ray flux, itself assumed to be proportional
 of the SN II rate. Thus, the 
 cumulated abundance of these light elements is proportional to the
 metallicity squared (secondary origin).
 Consequently, the observed linear behavior
 of Be and B is difficult to explain in terms of standard GCR (see also 
 Lemoine \etal 1997).

 A note of caution, however, since oxygen is the main progenitor of Be, the 
 apparent linear relationship between Be and Fe could be misleading if O
 is not strictly proportional to Fe. Indeed, new observations about the O/H vs
 [Fe/H] correlation (Israelian \etal 1998 and Boesgaard \etal 1998b) lead to a
 slope less than 1, contrary to previous studies. 
 So the Be-O relation would not be anymore linear but its slope
 would be about 1.5, leaving open the question of the primary and
  secondary origin
 of Be (Fields and Olive 1998a).
  Concerning Boron, as shown in Fields and Olive (1998a),  neutrino spallation 
  is necessary to fit  the B-Fe relationship.
 In this case, a large if not
 diverging B/Be ratio is predicted below [Fe/H] = -2.5, whereas observations
 seem to show a quasi constant ratio down to very low metallicity. Moreover,
 it has to be shown that this pure standard GCR solution  overcomes
 energetic 
 difficulties (Ramaty \etal 1996, 1997, 1998).

 Before ruling on models, 
we stress that there is
 a large dispersion in the [O/Fe] vs [Fe/H] observations. 
 Mc Williams (1997) shows clearly this dispersion in his figure 3, which
 is a compilation of the avalaible data until 1997, from [OI] results
 (Edvardsson \etal
 1993, Spite and Spite 1991, Barbuy 1988, Kraft \etal 1992, Sneden \etal 1991,
 Shetrone 1996), from the OI triplet (Abia and Rebolo 1989, Tomkin \etal 1992
 and from OH lines (Nissen \etal 1994, Bessell \etal 1991). Moreover, the 
[$\alpha$/
Fe] vs [Fe/H] where $\alpha$ = Mg, Si, Ca, S, Ti,
 (Cayrel 1996, Ryan, Norris and Beers 1996)
 show a plateau from about [Fe/H]= -4 to -1; on nucleosynthesis grounds, 
it would be surprising that
 oxygen would not follow this trend. Moreover, using all the published
 nucleosynthetic yields (Woosley and Weaver 1995, Thielemann, Nomoto and
 Hashimoto 1996) it seems impossible to fit the O/H vs [Fe/H] relation of
 Israelian \etal (1998) and Boesgaard \etal (1998b) since the required
 oxygen yields is unrealistic. 
To come back to observations, these two
 papers contradict a large number of former works on the variation of the O/Fe
 ratio with metallicity. Both are best on high quality observations, but depends
 upon the same set of physical assumptions: same empirical corrections of OH 
 oscillator strengths derived from solar spectrum (Balachandran and Bell 1998),
 use of the O I IR triplet, for which no agreement between theory and
 observation exist for the sun itself (Kiselman and Nordlund 1995),
 etc... So a definitive statement 
 on the variation of O/Fe with metallicity has also to explain why O/Fe ratios
 based on the [O I] line and Fe II lines, remarkably insensitive to departure
 to LTE and to the value of continuous opacity (Nissen and Edvardsson 1992) are
 wrong. Moreover, the situation could be even more complicated since 
 a  recent work (Thevenin and Idiard 1998) indicates that NLTE
 corrections for iron are needed for metal poor stars. 
 
 In the framework of the LEC model developed up to now, 
 related physically to the acceleration of SN II ejecta in the superbubbles
 produced by the OB associations (Parizot, 1998 and Parizot\etal 1997)
 , the linear relation between 
 Be, B vs [Fe/H] is naturally obtained, leading to a quasi constant B/Be ratio, 
 perhaps slightly affected by
  neutrino spallation ( Vangioni-Flam \etal 1996) that generates \b11 in
 agreement with the meteoritic \b11/\b10 ratio.
  \li6 is also produced by this same physical process. In the next section
 we will specifically study the evolution of \li6 in the light of the new 
 constraints set by its observation in Pop II stars. 

\section { Galactic evolution of \li6}

 At a 
given time, when a star forms, it inherits the composition of the 
ISM at this particular moment. Is the atmospheric Li abundance preserved in 
the course of stellar evolution?
Indeed, the depletion factor is still unknown in spite of 
many efforts to fix it (Chaboyer 1994, Vauclair and Charbonnel 
1995, Deliyannis \etal 1996, Chaboyer 1998, Pinsonneault \etal 1998, Cayrel
 \etal 1999).
In this context, it is necessary to compare the calculated and/or estimated 
\li6/H  in the ISM to the value measured in HD 84937 and BD +$26 \deg3578$ 
to get the \li6 
depletion factor at least in these stars, typical of all stars of the Spite's
 plateau. The \li6 abundance observed in their atmosphere 
  is necesserally equal or lower than that of the interstellar 
medium out of which these objects have formed since this isotope can 
only be depleted in stellar atmosphere due to proton capture. If 
there is no depletion in stars the observed points are located right 
on the evolutionary  curves.

Aimed at evaluating this \li6 depletion factor in stars, we propose two
 procedures, the first is based on a well founded evolutionary model
 of light elements and the second is based
 on an empirical criterion which is model independent.

 i) The Be and B evolution has been followed as a function of [Fe/H], 
 relying on a model invoking 
 fast nuclei ($\alpha$, C, O) originating from SN II within superbubbles,
 model S1
 (Bykov 1995, Parizot \etal 1997); here, only
 massive stars remain in the cavity (50 - 100 \msun) due to their
 short lifetime.
 They feed the cavity  with C and O nuclei which are
 injected by winds and explosions and are accelerated 
 in the low density gas by weak reflected shocks.
 In the early Galaxy, these extended acceleration sites
 would be sustained and filled
 essentially by SN II exploding in OB associations (Parizot \etal 1999).
 Later on,
 in the disk phase, Wolf Rayet stars would also participate, since the
 stellar winds intensify at increasing metallicities (Meynet \etal 1994). 

 The evolution of the \li6 abundance in the ISM as a function of time (or 
[Fe/H]) is calculated according to the same formalism and hypotheses than 
in Vangioni-Flam \etal (1998), and the progressive \li6 enrichment in the 
 ISM is followed together with that of Be and B.
 A step forward relative  to Vangioni-Flam \etal (1998) is to  take into
 account the new constraints given by \li6 and \be9 observations in the two
 considered stars. Indeed, the high \li6/\be9 ratio (20 - 80) at low metallicity
 (respective to
 the meteoritic
 value, about 6) leads us to consider a variation of the composition
 of the LEC. We
 invoke, to explain these preliminary observations, the fact that in the course 
 of the evolution of the Galaxy the composition of the superbubble supplied by
 the most massive stars is O rich at the beginning and  becomes 
 progressively carbon rich
 due to the increasing contribution of massive mass loosing stars
  (Maeder 1992, Portinari \etal 1998). Thus 
 the source composition of the LEC accelerated in these cavities changes as
 well, and in turn, the \li6/\be9 ratio produced by its spallation, as shown
 in figures 3 and 4.

  A grid of stellar yields calculated at different metallicities
 has been released by Woosley and Weaver (1995).  At very low metallicity,
 the composition of the U40B model of these  authors is taken as
 representative of
 the matter accelerated in superbubbles.
  This LEC composition is propagated and the resulting isotopic ratios of
 light species 
 are calculated (Parizot \etal 1999 in preparation). As the ISM metallicity 
 increases, the source composition is taken variable. We adopt  a spectrum
 of the form : N (E) dE  = k$E^{-1.5}$ exp (-E/Eo), with Eo = 30 MeV/n as
 previously (Vangioni-Flam \etal 1998). In
 figure 3, the behaviour of the \li6/\be9 ratio versus C/O for three  He/O
 ratios is shown.  It is interesting to note that the He/O ratio in the
 ejecta  and winds of massive stars varies much less than the C/O ratio in
 the course
 of the evolution.
 We see that as C/O increases the evolution going on,
 \li6/\be9 decreases from  about 20 to 6 (for He/O = 10).
 For comparison, we show
 in figure 4 the same diagram but with Eo = 10 MeV/n; the \li6/\be9 ratio
  decreases more steeply starting from about 70 until 6.
  The related light element yields for Eo= 30 MeV/n
 at low C/O and at He/O = 10  corresponding to the early Galaxy,
   are in better agreement with the observational constraints.
 
 These diagrams allow to derive the \li6/\be9 ratio produced by the LEC primary
 component for any composition and metallicity at any time. As a consequence, 
 it is clear that in the disk phase this ratio will decrease,
  meaning that the 
 production of \li6 per \be9 nucleus diminishes. So, to cover the \li6
 disk evolution, the source composition has to be taken variable vs time,
  taking into account in details the variation of He, C, O in the source
 composition (Parizot \etal 1999).
 In this work, as a first approach, we include this variation, reducing the 
 \li6 yield progressively (figure 3). 
 Typically,  in relation with U40B (Woosley and Weaver 1995) :  Li/Be =60,
 B/Be = 19, \li6/\be9 =20 ending at present with \li6/\be9 at about 7.
 Note that the higher value of this last ratio 
obtained with Eo= 10 MeV/n at low C/O, (about 70) corresponds
 to the upper observational 
 value (Hobbs and Thorburn 1994). However, this extremely high Li production
     is quite inconsistent with other constraints, specifically with
 the global behaviour of Be. In this case, the model leads to Li/Be = 271 
 in the early Galaxy, which is too high. 
    
The yields obtained with the truncated spectrum described above are injected
 in the galactic evolutionary model. We remind that the mass range of $\alpha$ ,
 C, O progenitors
 is 50 - 100 \msun (Model S1) as implied by the superbubble scenario. For 
 comparison we have extended the mass domain (10-100 \msun, Model S2) to see
 the effect of the enlargement of the stellar population involved as
 in Vangioni-Flam \etal (1998) using the Woosley and Weaver (1995) yields at 
 low metallicity; this S2 model runs
 with a constant source composition contrary to S1 model since the dominant
 stars in this
 hypothesis have about 15 \msun, and in this case, they are not affected
 by mass loss at any metallicity.

Figure 2 presents the evolution of \li6/H versus [Fe/H] for i) standard
 GCR ii) LEC from superbubbles for two mass ranges.
 The full line  shows the behaviour of \li6  for model S1 (taking into account
 the variation of the source composition). The dotted line represents
  model S2.
 The  HD 84937 and BD +$26 \deg3578$ measurements  are plotted for
 confrontation. 
Though there are only two  observational points, it is sufficient to draw 
important conclusions thanks to the good knowledge of the 
evolutionary process at work.
As can be seen, at the metallicity of these stars, 
 the quasi absence of \li6 depletion is demonstrated since models 
 S1 and S2 go through the observational data.
   Note that standard GCR (Figure 2, dash-dotted line) assumed to have the same 
energy spectrum than observed presently (i.e. no strong excess at low energy, 
 Lemoine \etal 1998) 
 plays an unsignificant 
role in the early evolution of \li6 and that cosmic ray alone 
cannot explain the amount of \li6 measured in these stars. This is not
  surprising 
 because, as seen previously, the LEC production is much stronger than the
 GCR one (by a factor of about 50). Fields and Olive (1998b), on their side, 
 fit the \li6 evolution taking into account, as mentionned above, the new
 O-Fe relation of Israelian \etal (1998). In this context, considering 
 exclusively the standard GCR component, the slope of the relation \li6/H vs
 [Fe/H] goes from 1 to about 0.6 leading to a good fit of all \li6
 observations. In our case, the variation of the LEC component induces 
 a flattening of the \li6 evolution in the disk phase, leading to the
 solar value.  

 At very low metallicity, [Fe/H = -4, which corresponds to the onset of
 the star formation in the Galaxy, models S1 and S2
 differ substantially, for example by
  about 1 dex at [Fe/H] = -4. These two cases lead to a range of very early
  \li6 abundance of 5.6 $10^{-14} $ to 3. $10^{-13}$, to be compared with the BBN
 calculations.
It is interesting to see that the theoretical description of the \li6 evolution
 in the early galaxy on the basis of the fast nuclei spallation  is essentially
 consistent within a factor of 2 with the observations of the two stars,
 which is indicative of low destruction of \li6 , if any.

ii) In a second approach, independently of the galactic evolutionary 
model, one can derive an 
absolute upper limit of \li6 depletion in HD 84937 and BD +$26 \deg3578$,
 using the 
following arguments 
a) the evolutionary curve of \li6 versus [Fe/H] is linear at least down 
 [Fe/H] = -3,
b) the maximum value of the Spite's plateau ending at about  
[Fe/H=-1.3] is  2.5$10^{-10}$ (see figure 2, horizontal thick
 line) (Spite
1997). Above this metallicity, which corresponds to the transition 
between the halo phase and the disk one, the Li abundance is rising, 
reaching 2. $10^{-9}$ at solar metallicity (Anders and Grevesse 1989).
 Thus, typically, 
spallative Li (thick full line, slope 1) cannot be higher than 2.5 $10^{-10}$
 at [Fe/H] = -1.3 to avoid 
crossing the Spite 's plateau which corresponds to a maximum 
\li6/H = $10^{-10}$, since almost all spallation models lead to 
 \li7/\li6 of about 1.5. This is an absolute upper 
limit of the \li6 production in the halo of the Galaxy. Indeed, the same
 kind of argument has been used to limit the theoretical yields of \li7 
 synthesized by neutrino spallation in SN II, in the early Galaxy (
Vangioni-Flam \etal 1996). The deduced maximum \li6 curve ( Figure 2,
 long and short dashed thick line)
 passes again across the error bar of the 
observed points. This indicates, independently of any model, that 
\li6 is essentially intact in the envelope of HD84937 and BD +$26 \deg3578$,
 provided 
 the \li6/H vs [Fe/H] evolution is linear. 
 Boesgaard \etal (1998a) advocate a 
larger dispersion of the  Spite's plateau (full range of a factor of 3); 
in this case, the maximum depletion factor of \li6 is about 2.

 We can conclude that the \li6 observed in HD84937 and BD +$26 \deg3578$ 
is in good agreement to
 the values expected from low energy spallation production; thus, there
 is little room
 for stellar destruction. Hence, \li7 which is more solid than \li6, is
 even less depleted. This result, besides the flatness and the small
 dispersion of the plateau is a new strong argument to consider the average 
 Li/H of the Spite's plateau as the primordial one.

 Finally,  for a primordial \li7 value of  1.6 $10^{-10}$
 we deduce \li6/H = 3-6 $10^{-14}$ which is consistent with the range deduced
 by both empirical and theoretical methods. 
This corresponds to $\eta$ = (1.8 to 3.8) $10^{-10}$ .
 These results are in agreement with other independent studies
 (Fields \etal 1996).
 It is gratifying to see that the \li6 evolution can be followed consistently
 from the standard Big Bang up to now. Thus, a pregalactic production
 through the $\alpha$ + $\alpha$ reaction is not required.

\section{Conclusion}

Thanks to the refined observations (Cayrel \etal 1998, Smith \etal
 1998)  of 
the \li6/\li7 ratio in two halo stars HD 84937, BD +$26 \deg3578$ 
 and considering the galactic 
evolution of \li6 driven by spallation, we arrive to the 
following conclusions :
i) \li6 and \li7 are not significantly depleted in the 
atmosphere of halo stars, consequently, this is a new
 and strong argument in favor of the primordial status of the Spite's plateau.
ii) the model of Lithium-Beryllium-Boron evolution 
induced by fast nuclei from SN II distinct from the 
classical GCR is consistent, since we can explain the whole evolution of \li6
 from BB to now including the Be, B evolution. However, another
 alternative has been
 developed on the basis of a new O/H vs Fe/H relationship (Fields and Olive
 1998a,b) who propose a solution based on standard GCR alone.
 Moreover, the new observational 
 constraints on \li6/\be9 ratio in Pop II stars have been integrated. 
 The fit of  all the data on the light elements leads to adopt a
 \li6/\be9 ratio of about 20 in the early galactic evolution.
  A consistent scenario has been proposed
 on the basis of shock acceleration of nuclei in superbubbles the composition
 of which is oxygen rich in the halo phase and becomes progressively carbon rich
 iii)  it is possible to 
 set a range (5.6 $10^{-14}$ to 3. $10^{-13}$) of the \li6 abundance
 in the very early galaxy, which can be
 compared favorably to the \li6 from BBN. Consequently,
 a pregalactic $\alpha$ + $\alpha$
 production of \li6 is not required.
iv) Considering both Li isotopes we get 1.8 $10^{-10}$ < $\eta$< 3.8
$10^{-10}$ in agreement with independent estimates based on 
deuterium and helium-4.
In conclusion, the addition of \li6 to the other light isotopes allows to
 confort the general picture of the cosmological nucleosynthesis followed
 by spallation processes.

Aknowledgements : we are endebted to Jurgen Kiener for his expertise on BBN 
nuclear 
 reaction rates we thank also warmly Etienne Parizot for providing us with the
 required spallative yields. We are grateful to Keith Olive
 and Brian Fields and Martin Lemoine for fruitful discussions.
 This work was supported in part
 by the  PICS 319, CNRS at the Institut d'Astrophysique de Paris.

\beginapjbib

\bibitem Abia, C., Rebolo, R., 1989, ApJ, 347, 186
\bibitem Anders, E. and Grevesse, N. , 1989, Geochim. Cosmochem. Acta,
 53, 197 
\bibitem Angulo, C. \etal 1998, At. Nucl. Data Tables, submitted
\bibitem Balachandran. S.C. and Bell, R.A. 1998, Nature, 392, 791
\bibitem Barbuy, B., 1988, AA 191, 121
\bibitem Bessell, M.S., Sutherland, R.S., Ruan, K., 1991, ApJL, L71 
\bibitem Boesgaard, A.M. and King, J.R. 1993, AJ, 106, 2309
\bibitem Boesgaard, A. M., Deliyannis, C.P., Stephens, A. \& King, J.R.
 1998, ApJ, 493, 206
\bibitem Boesgaard, A.M., King, J.R., Deliyannis, C.P.
  and Vogt, S.S. 1998b, submitted 
\bibitem Bonifacio, P. \& Molaro, P. 1997, MNRAS, 285, 847
\bibitem Bykov, A.M., 1995, Space Sci.Rev. 74, 397
\bibitem Caughlan, G.R. \& Fowler, W.A. 1988, At.Nucl.Data Tables, 40, 283
\bibitem Cayrel, R. 1996, A.A R Rv, 7, 217
\bibitem Cayrel, R., Spite M., Spite F., Vangioni-Flam, E., Cass\'e,
 M. and Audouze, J. 1998, AA submitted
\bibitem Cayrel, R.,Lebreton, Y., Morel, P. 1999, to be published in 
 "Galaxy evolution : connecting the distant Universe with the local fossil
 record", Meudon, sept 1998, Edts M.Spite \etal, Kluwer, Dordrecht 
\bibitem Cecil, F.E., Yan, J. \& Galovich, C.S. 1996, Phys.Rev.C., 53, 1967
\bibitem Chaboyer, B. 1994, ApJ, 432, L47
\bibitem Chaboyer, B. 1998, submitted,  astroph/9803106
\bibitem Charbonnel, C., Vauclair, S. \& Zahn, J.P. 1992, AA, 255, 191
\bibitem Delbourgo-Salvador, P. \& Vangioni-Flam, E. 1992, in " Origin and
 Evolution of Elements", Edts Prantzos \etal, Cambridge University Press, p. 52
\bibitem Deliyannis, C.P., Demarque, C. \& Kawaler, S.D. 1990, ApJS 73, 21
\bibitem Deliyannis, C.P., King, J.R. and Boesgaard, A.M., 1996, 
BAAS, 28, 916
\bibitem Duncan, D.K., Lambert, D. \& Lemke, M. 1992, ApJ 401, 584
\bibitem Duncan, D.K. \etal 1997, ApJ, 488, 338
\bibitem Edvardsson, B.\etal, 1993, AA, 275, 101
\bibitem Fields, B.D., Kainulainen, K., Olive, K.A. \& Thomas, D. 1996, 
New Astronomy, 1, 77
\bibitem Fields, B., Olive, K. 1998a, submitted
\bibitem Field, B., Olive, K. 1998b, submitted
\bibitem Garc\'ia-L\'opez \etal 1998, ApJ, in press
\bibitem Kiener, J. \etal 1991, Phys.Rev.C., 44, 2195
\bibitem Kraft, R.P., Sneden, C., langer, G.E., Prossen, C.F., 1992,
 Astron. J., 104, 645 
\bibitem Hobbs, L.M. \& Thorburn, J.A. 1994, ApJ, 428, L25
\bibitem Hobbs, L.M. \& Thorburn, J.A. 1997, ApJ, 491, 772
\bibitem Israelian, G., Garc\'ia-L\'opez, R.J., Rebolo, R. 1998, ApJ, to be
 published
\bibitem Kiselman, D. and Nordlund, A. 1995, A\&A 302, 578
\bibitem Lemoine, M., Schramm, D.N., Truran, J.W. and Copi, C.J. 
1997, ApJ, 478, 554
\bibitem Lemoine, M., Vangioni-Flam, E. \& Cass\'e, M. 1998, ApJ, 499, 735
\bibitem Mc Williams, A., 1997, Ann.Rev.Astron.Astrophys., 55, 503
\bibitem Maeder, A., 1992, AA, 264, 105
\bibitem Meynet, G. \etal 1994, AAS, 103, 97
\bibitem Mohr, P. \etal 1994, Phys.Rev.C., 50, 1543
\bibitem Molaro, P., Primas, F., \&  Bonifacio, P. 1995, A \& A,
295, L47 
\bibitem Molaro, P., Bonifacio, P., Castelli, F. and Pasquini, L. 1997,
 AA 319, 593
\bibitem Nissen, P.E., Edvardsson, B. 1992, A\&A 261, 255
\bibitem Nissen, P.E., Gustafsson, B., Edvardsson, B., Gilmore, G., 1994
  AA 285, 440
\bibitem Nollet, K.M., Lemoine, M. \& Schramm, D.N. 1998, Phys. Rev. C56, 1144
\bibitem Parizot, E., Cass\'e, M., Vangioni-Flam, E. 1997, AA 328, 107
\bibitem Parizot, E. 1998, A\&A in press
\bibitem Parizot \etal 1999, in preparation
\bibitem Pinsonneault, M.H., Deliyannis, C.P. \&
 Demarque, P. 1992, ApJS, 78, 181
\bibitem Pinsonneault, M.H., Walker, T.P., Steigman, G. \& Naranyanan, V.K. 1998
 , ApJ submitted
\bibitem Portinari, L., Chiosi, C. \& Bressan, A. 1998, A\&A 334, 505
\bibitem Rauscher, T. \& Raimann, G. 1997, ref nucl-th/9602029
\bibitem Ramaty, R., Kozlovsky, B. \& Lingenfelter, R.E. 1996, ApJ 
456, 525
\bibitem Ramaty, R., Kozlovsky, B., Lingenfelter, R.E. \& Reeves, H. 1997, ApJ
 488,730
\bibitem Ramaty, R., Kozlovsky, B. \& Lingenfelter, R.E. 1998, Physics Today,
 51, no 4, 30
\bibitem Ryan, S.G., Norris, J.E. \& Beers, T.C. 1996, ApJ 471, 254
\bibitem Read, S.M. \& Viola, V.E. 1984, Atomic Data and Nuclear Data Tables,
 31, 359
\bibitem Shetrone, M.D., 1996, Astron.J., 112, 1517
\bibitem Schramm, D.N 1992, in "Origin and Evolution of the Elements", Edts
 Prantzos \etal , Cambridge University Press, p. 112
\bibitem Schramm, D.N 1994, in "Light Element Abundances", Edts P. Crane, ESO
 Astrophysics Symposia, Springer, p. 50
\bibitem Schramm, D.N. and Turner, M. 1998, Rev,Mod.Phys. 70,303
\bibitem Smith, V.V., Lambert D.L. \& Nissen P.E. 1993, ApJ, 408, 262
\bibitem Smith, V.V., Lambert, D.L. and Nissen, P.E., 1998, ApJ, 506, 405
\bibitem Sneden, C., Kraft, R.P., Langer, G.E., 1991, Astron,J, 102,2001
\bibitem Spite, M., Spite, F., 1991, AA 252, 689
\bibitem Spite, F. \& Spite M. 1982, AA, 115, 357
\bibitem Spite, M. 1997, in "Fundamental stellar properties: the Interaction
between observation and theory", Sympos. IAU,Edts T.R. Bedding \etal, Kluwer,
  p. 185
\bibitem Thevenin, F. \& Idiard, J. 1998, XXX in press
\bibitem Thielemann F.K, Nomoto, K. \& Hashimoto, M. 1996, ApJ, 460, 408
\bibitem Thomas, D., Schramm, D.N., Olive, K.A. \& Fields, B.D., 
1993, ApJ, 406, 569 
\bibitem Tomkin, J., Lemke, M., Lambert, D.L., Sneden, C., 1992, Astron.J, 104
 1568
\bibitem Vangioni-Flam E., Cass\'e, M., Olive, K. \& Fields B.D. 
1996, ApJ 468, 199
\bibitem Vangioni-Flam, E., Cass\'e, M. and Ramaty, R. 1997, in 
Proceedings of the 2nd INTEGRAL Workshop, "The Transparent 
Universe", Saint Malo, France, ESA, p. 123
\bibitem Vangioni-Flam, E. Ramaty, R., Olive, K.A. \& Cass\'e, M. 1998, AA, 
 accepted
\bibitem Vauclair, S. \& Charbonnel, C. 1995, AA, 295, 715
\bibitem Vauclair, S. \& Charbonnel, C.1998, ApJ, 502, 372
\bibitem Woosley, S.E. \& Weaver, T.A. 1995, ApJS, 101, 181 
\endapjbib

\section*{Figure Captions}

{\bf Figure 1:}

BBN of \li6, \li7, \be9, \b10, \b11. The abundance of light isotopes by 
 number is presented versus the baryon/photon ratio (the $\eta$ parameter).

 For 
the reaction \b10 (p,$\alpha$) \be7,
  two different rates have been used. Full line 
(Caughlan and Fowler 1988) and dashed line (Rauscher and Raimann 1997).

 {\bf Figure 2:}

Rise of the \li6 abundance  from the birth
of the Galaxy  up to now. 

The logarithm of the abundance by number of lithium 
isotopes is shown against the [Fe/H] = (log(Fe/H)*/log(Fe/H) \msun).
 HD 84937 observation is from Cayrel \etal (1998) and BD +$26 \deg3578$ is from
 Smith \etal (1998) (squares). Error bars are taken as 0.2 dex for [Fe/H], 
 (+0.23, -0.19)dex  for \li6/H of HD 84937, and 0.2dex for \li6/H of BD +$26
 \deg3578$.
 The corresponding Li/H for these stars are indicated by triangles.
 The meteoritic value is from Anders and Grevesse (1989).

 full line : galactic evolution of \li6 for model S1 (mass range : 50 -
 100 \msun)
 
 dotted line : galactic evolution of \li6 for model S2 (10-100 \msun)

 dot-dashed line : galactic evolution of \li6 for standard GCR.

 Thick lines :

 1. horizontal full line : upper limit of the Spite's plateau  (Li/H)

 2. Slope 1 full line : upper limit of the spallative \li7 plus \li6
 evolution   consistent with the upper limit of the Spite's plateau.

 3. Slope 1 dashed line : maximum  \li6 evolution deduced from previous curve
 (knowing that \li7/\li6 = 1.5 in spallative processes).

{\bf Figure 3:}

 The \li6/\be9 ratio as a function of C/O for He/O =1, 10, 50 for Eo = 30 MeV/n.

  The source abundances are from model U40B (Woosley and Weaver 1995).

{\bf Figure 4:}

 Same as figure 3 but for $E_{0}$ = 10 MeV/n.

 
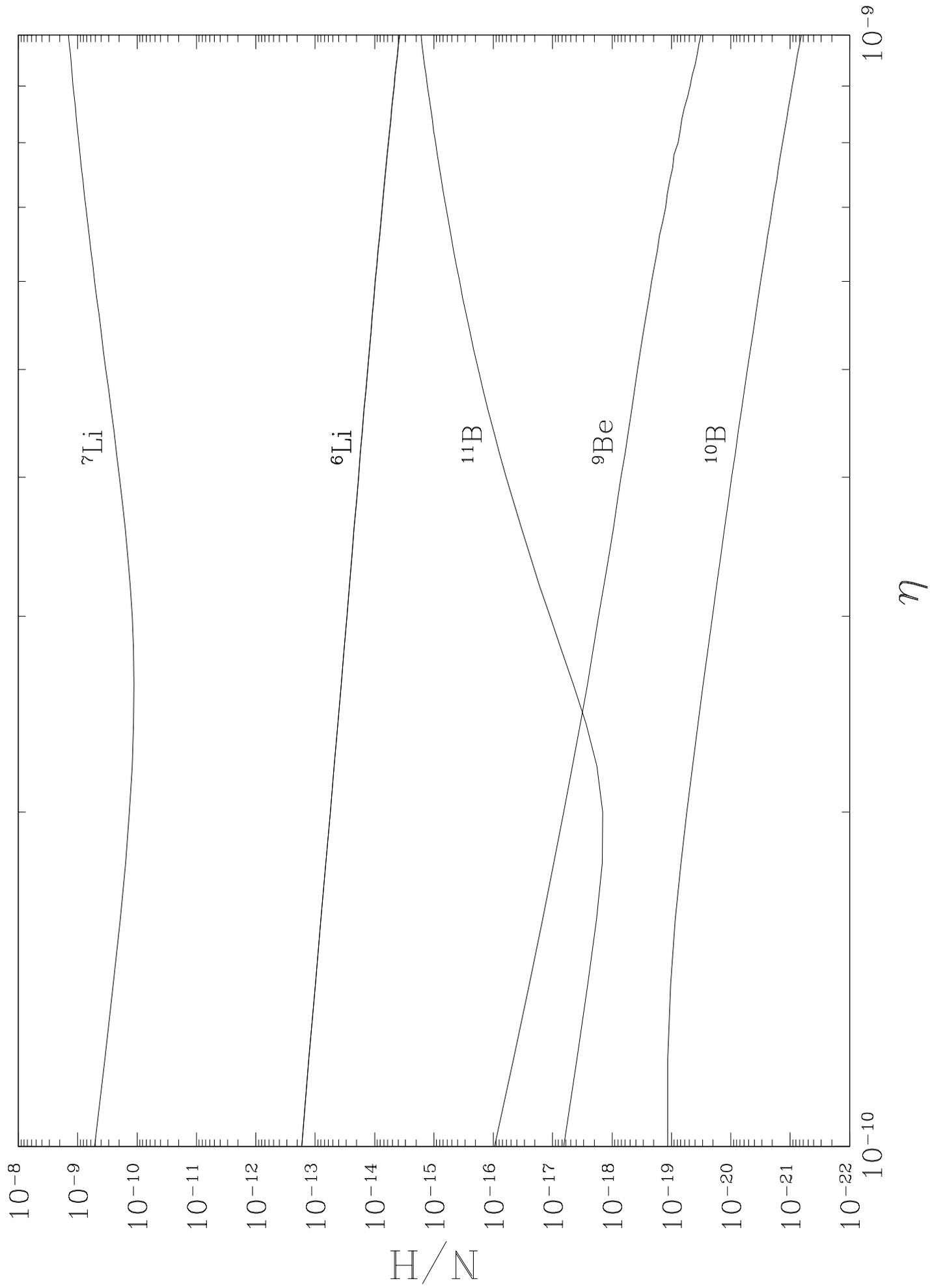
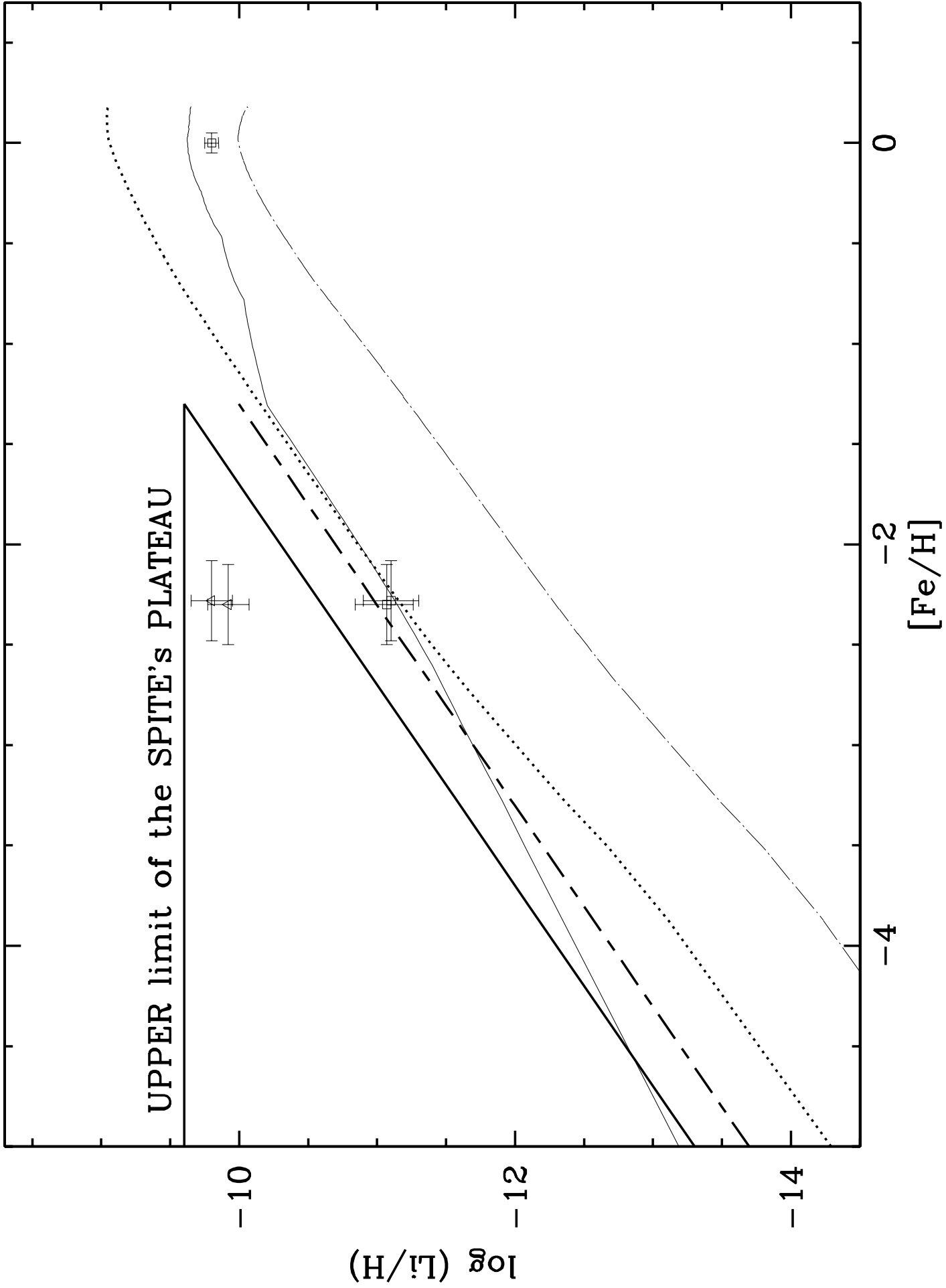
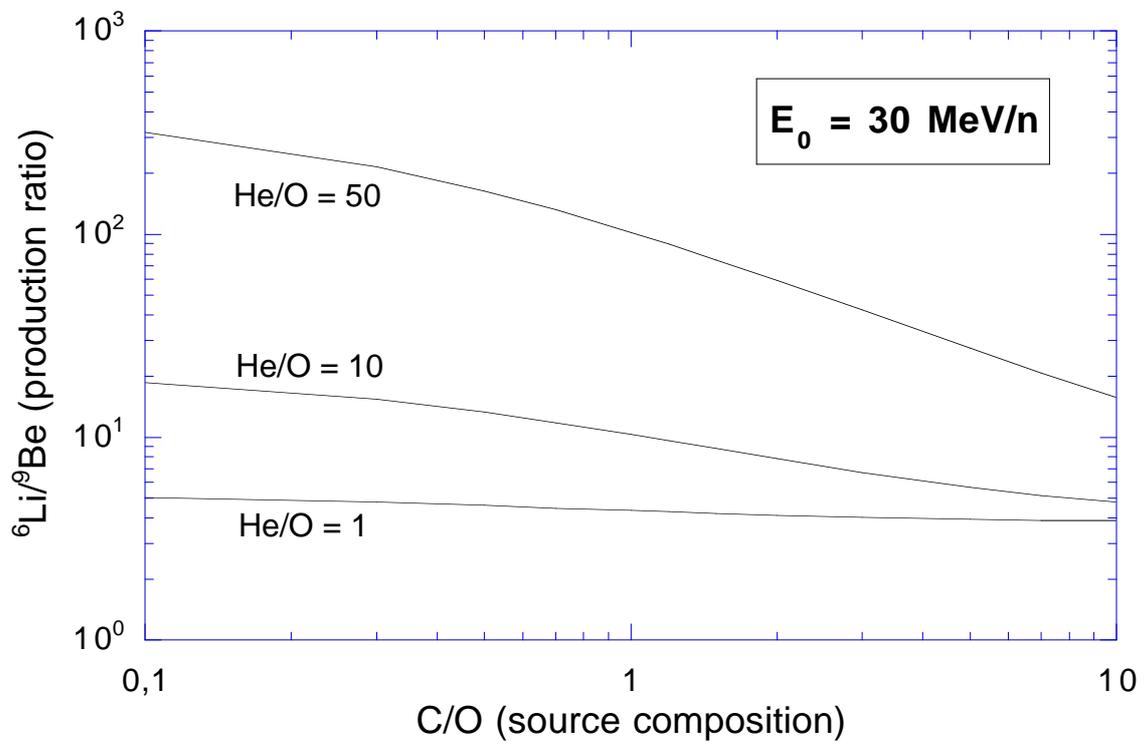
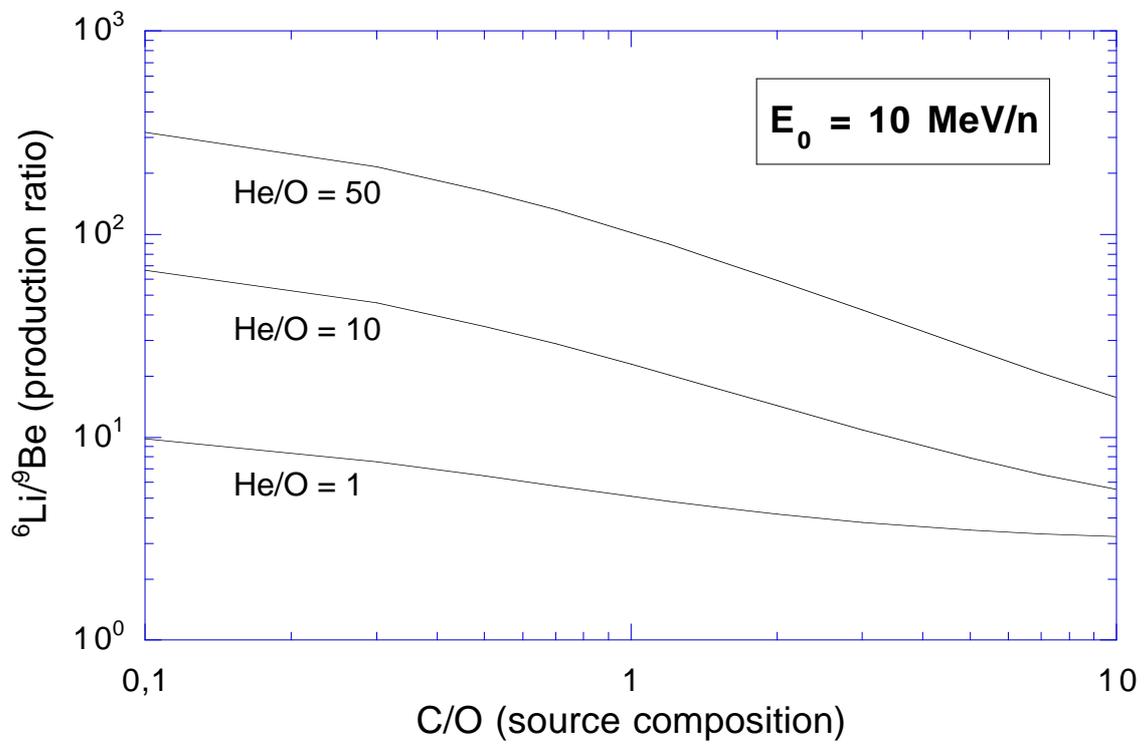
\end{document}